\documentclass[twocolumn,aps,prl,groupedaddress,amssymb,showpacs]{revtex4}
\usepackage{graphicx}
\usepackage{dcolumn}
\usepackage{amsmath}
\usepackage{bm}

\begin{document}
\title{Compton light pressure and spectral imprint of relic radiation on cosmic electrons}
\author{A. E. Kaplan} 
\affiliation{Electr. and Comp. Engineering Dept,
The Johns Hopkins University, Baltimore, MD 21218 } 
\email{alexander.kaplan@jhu.edu}

\date{\today}
\begin{abstract}
A fully QED/relativistic theory of light pressure 
of CMB radiation and Fokker-Planck
equation for electron distribution combined with
cosmologic relation for CMB temperature, $T$,
yields analytic results for the evolution
of the distribution over large span of time and energies. 
A strong imprint of CMB on electrons transpires $via$ formation
of ``frozen non-equilibrium'' state of
electrons in current epoch, and possible
existence of cutoff and narrow spectral lines
as remnants of high-$T$ sources.
\end{abstract}
\preprint{Submitted to \emph{Phys.~Rev.~Lett.}}
\maketitle
Beginning with its discovery [1], 
light pressure, in particular
that by an isotropic
Cosmic Microwave Background (CMB) [2]
on charged particles,
resulting in the loss of their momentum/energy [3]
$via$ inverse Compton scattering [4],
played a substantial 
role in astrophysics and cosmology.
It strongly affects high-energy baryons' fast decay
facilitated by the pion production 
and high-energy photons 
$via$ secondary production of virtual 
pairs, imposing an upper limit [5]
on cosmic rays energy.
One would expect even stronger 
CMB interaction  with electrons,
since due to their fundamental nature,
their momentum loss
can be treated more thoroughly
using the QED theory [6] 
of photon-electron scattering.

In this Letter, we show that very interesting effects
in such an interaction
can be elicited at both low- and high-energy domains.
At low-energy, we predicts the
formation of a ``frozen non-equilibrium'' state
of electrons as the universe expands (including present epoch):
due to its cooling, CMB fails to enforce 
thermal equilibrium on electrons,
and let them keep constant temperature forever ($10 - 20 K$).
At high-energies ($> E_C \sim 10^{14} eV$)
we predict transformation of
initial thermal electron spectra into
narrow lines followed by a cutoff near $ E_C$.
We limit our consideration here
only to the momentum decay due to CMB
and do not consider other evolution channels
(such as e. g. synchrotron radiation due to galactic magnetic
fields, secondary effects due to decay of protons,
anisotropy fluctuations due to Sachs-Wolfe effect, etc).

\textbf {\emph {Toolkit}}. 
A major tool here
is a light pressure $F$ by a black-body isotropic
radiation (in particular, CMB) on an electron.
In [7] we derived a general
QED/relativistic formula for $F$
based on Lorentz transformation of
an $arbitrary$ spectrum $\rho ( \epsilon )$
of (dimensionless) photon energies 
$\epsilon = \hbar \omega / m_0 c^2$
(here $m_0$ is the rest mass of an electron),
in a frame, $L$, where the radiation is isotropic,
upon transition from that frames
to the frame, $R$, where the particle is at rest.
The theory is valid also for any energy
dependence of a cross-section $\sigma (  \epsilon )$
of scattering of an $\omega$-photon at a particle.
For a particular case of 

(1) a black-body (Planck) radiation
of an arbitrary temperature $T$,
with its spectral, $\rho_{_{BB}}$,
and total energy,
$W_{_{BB}} = W_{_C} \int_0^\infty 
\epsilon \rho_{_{BB}} ( \epsilon ) d \epsilon$
densities in the $L$-frame  being
\vspace {-.2in}
\begin{equation}
\rho_{_{BB}} ( \epsilon ) d \epsilon  = \frac { 8 \pi \epsilon^2 d \epsilon }
{e^{ \epsilon / \theta }  - 1};
\ \ \ \ \
W_{_{BB}} = \frac{8 \pi^5}{15} W_{_C} \theta^4
\tag{1}
\vspace {-.05in}
\end{equation}
where $\theta = k_{_B} T /  m_{_0} c^2$
is a dimensionless temperature
(with $T = T_0 \approx 2.725 K$, and
$\theta = \theta_0 \approx 0.534 \times
10^{-9}$ for present CMB),
$k_{_B}$ is the Boltzmann constant,
$W_{_C} = m_{0} c^2  /  \lambda_{_C}^3 $
is a ``Compton energy density'', and
$\lambda_{_C} = 2 \pi \hbar / m_0 c$
is the Compton wavelength, and 

(2) an electron as a scattering particle,
with its cross-section, $\sigma ( \epsilon )$,
described by Klein-Nishina theory
[6] accounting for virtual electron-positron
pair creation/annihilation in the 1-st order
of $\alpha = e^2 / m_0 c^2 \approx 1/137$,
for an $any$ $\epsilon$,
so $\sigma \approx$ $ \sigma_0 = 
( 8 \pi / 3 ) r_0^2$ at $\epsilon \ll 1$
is the Thompson cross-section of electron,
where $r_0 = e^2 / m_0 c^2$ is the
classical electron EM-radius,
and $\sigma \approx \sigma_0 ( 3 / 8 \epsilon )
[ ln ( 2 \epsilon ) + 1/2 ]$ at $\epsilon \gg 1$
in a Compton domain,
we found a simple and precise 
analytic approximation [7] for the dimensionless light
pressure force $f = F t_C / m_0 c$ in
terms of the electron momentum $\mu \equiv p / m_0 c$,
relativistic factor $\gamma = \sqrt { 1 + \mu^2 }$,
and temperature $\theta$ as
\vspace {-.05in}
\begin{equation}
f ( \mu , \theta ) = \frac {d \mu} {dt} t_C
\approx - \frac {\mu \theta^3 } q ln ( 1 + K_C ) ;
\ \ \ \ \ K_C = \gamma \theta q
\tag{2}
\vspace {-.05in}
\end{equation}
where $K_C$ is a ``Compton factor'',
$q = 10.0$ is a numerical fitting parameter,
and $t_C$ is a ``Compton time scale'':
\begin{equation}
t_{_C} = 135 \lambda_{_C} / 64 \pi ^4  \alpha^2 c  
\approx 3.2515 \times 10^{-18} s ; \ \ \ \
t_{_C} \propto \hbar^3
\tag{3}
\end{equation}
Eq. (2) remains true in the entire span of momenta, $\mu \in ( 0 , \mu_{Pl} ) $,
where $\mu_{Pl} = k_B T_{Pl} / m_0 c^2 \approx 2.4$ $ \times 10^{22}$
is the highest momentum in the universe
related to the Planck temperature, $T_{Pl} \approx 1.417 \times 10^{32} K$.
The Thompson domain corresponds to
$K_C \ll 1$ (hence $\theta \ll 1$), with
\vspace {-.15in}
\begin{equation}
f \approx - \mu \gamma \theta^4 ,
\ \ \ \ \ \gamma = \sqrt{ 1 + \mu^2 }
\tag{4}
\end{equation}
consistent with a well known result (see e. g. [8]);
note that it is still good for relativistic case, 
$| \mu | \sim  \gamma \gg 1$, as long as $| \mu | \ll \theta^{-1}$.
The Compton (QED) domain is defined by $K_C \gg 1$, 
and its threshold, $K_C = 1$, for present CMB
corresponds to the energy $E_C \sim 10^{14} eV$.

While Eq. (2) can be directly
used to calculate the decay 
of momentum $\mu (t)$ for a given
initial condition (see below),
the temporal evolution of electron $distribution$
should be found from a Fokker-Planck equation
for the diffusion in the momentum space [9].
We define a distribution function, $g^{(e)} (\mu , t )$
of electrons as the number of electrons per
elements of solid angle $d O$, momentum, $d \mu$,
within a unity of coordinate space, and a density number,
$\rho^{(e)} (\mu , t ) = 4 \pi \mu^2 g^{(e)}$,
and note that in the expanding space/universe, 
we need  to use an also expanding unity of coordinate space.
Assuming then that (a) the electron distribution
is isotropic, same as CMB, (b) the total number
of electrons is invariant,
$\int_0^{\infty} \rho^{(e)} d \mu = inv$,
and (c) the thermal
equilibrium of a relativistic gas at any $\theta = const$ is due to the
Maxwell-J\"{u}ttner (MJ) distribution [10],
\begin{equation}
g_{_{MJ}}^{(e)} \propto  e^{- \gamma /\theta}
[\theta K_2 ( 1 / \theta ) ]^{-1}
\tag{5}
\end{equation}
where $K_2$ is the modified Bessel function of the second order,
with MJ being a relativistic generalization
of the Maxwell-Boltzmann (MB) distribution,
\vspace {-.05in}
\begin{equation}
g_{_{MB}}^{(e)} \propto  e^{- \mu^2 / 2 \theta} \theta^{- 3/2}
\tag{6}
\vspace {-.05in}
\end{equation}
we found [7] a Fokker-Planck equation
for $g^{(e)} ( \mu , t )$, as
\vspace{-.1in}
\begin{equation}
\frac {\mu^2 \partial [ g^{(e)}] } {\partial ( t / t_C ) } +
\frac {\partial } {\partial \mu}
\left\{ \mu^2 f ( \mu , t ) \left[ g^{(e)} +
\theta ( t ) \frac {\gamma } {\mu}
\frac {\partial g^{(e)}} {\partial \mu} \right] \right\} = 0
\tag{7}
\end{equation}
In non-relativistic case 
[$\gamma \approx 1$ in Eq. (4)], it comes to
\vspace{-.05in}
\begin{equation}
\frac {\mu^2 \partial [ g^{(e)}] } {\partial ( t / t_C )} =
\theta^4 ( t ) \frac {\partial } {\partial \mu}
\left\{ \mu^3 \left[ g^{(e)} +
\frac {\theta ( t )} {\mu}
\frac {\partial g^{(e)}} {\partial \mu} \right] \right\} 
\tag{8}
\end{equation}
Finally, when tackling the dynamics of CMB temperature, $T$,
due to universe expansion, we recall that 
it is related to the redshift $z$ as $T (t)  / T_0 = 1 + z (t)$
($T_0$ is a present value), and thus is governed
by a standard cosmologic relation [11,12]:
\begin{equation}
d T / dt = - T H_0 
[ \Omega_{\Lambda} + \Omega_M ( T / T_0 )^3 + \Omega_R ( T / T_0 )^4 ]^{1/2}
\tag{9}
\end{equation}
where $H_0$ is a present Hubble constant
(with $H_0^{-1} \approx 4.414 \times 10^{17} s$
being an approximate age of the universe),
$\Omega$'s are the fractions of respective
forms of energy in critical energy density
(it is a common convention that our universe
is flat, hence $ \Omega_{\Lambda} + \Omega_M + \Omega_R = 1$) 
with commonly accepted values
$\Omega_{\Lambda} \approx 0.7$
(a vacuum energy density fraction,
or cosmological (or $\emph {dark energy}$) constant, 
a major contributor to the current rate of the universe expansion),
and $\Omega_R \sim 0.85 \times 10^{-4}$ -- radiation, or relativistic
fraction, dominant at the earlier stage
of the universe, and non-relativistic, or ``matter''
fraction $\Omega_M  \approx 0.3$.

{\bf {\emph {Frozen non-equilibrium}}}.
How promptly an electron distribution 
equilibrates with changing CMB temperature, $\theta (t)$?
To find this out,
we compare time scales of both of them.
That of CMB is roughly the age of universe, $t_U$, at
a given CMB temperature, $\theta$, i. e. 
$t_U (\theta ) = \int_{\infty}^{\theta} d \theta / ( d \theta / dt )$.
Using Eq. (2), we evaluate the time scale as
inverse momentum decay rate,
$t_{\mu} = \mu / | d \mu / dt |$,
at the peak of distribution $\rho^{(e)} ( \mu , t )$
for a given equilibrium, 
and then solve the equation $t_U (\theta ) = t_{\mu} ( \theta )$
for a split-point $\theta = \theta_{spl}$ numerically.  
With $\Omega_M >> \Omega_R$, we found then that
$T_{spl} / T_0 \lesssim 10$, 
which is consistent with 
detailed calculations, Fig. 2,
and the split occurred at $ t \lesssim 0.05 H_0^{-1}$.
The main point here is that
it falls far within Thompson domain, $\theta_{spl} \lesssim 0.5 \times 10^{-8} \ll 1$,
so that the electron kinetics could be described by classical Eq. (8).

In the earlier epoch, the thermalization
of electrons happened almost instantaneously,
so that their distribution is
described by Eq. (5) and (6)
with the temperature, $\theta_e (t)$ of this distribution
following almost exactly the CMB temperature, $\theta (t)$.
To investigate what happened after 
they start diverging near $\theta = \theta_{spl}$,
we need to solve Eq. (8) with an initial 
condition given by MB-distribution (6)
at any point $1 \gg \theta \gg \theta_{spl}$.
Most luckily, that partial derivative equation happens to
have an exponential MB-distribution Eq. (6) as an self-similar
solution [13], where the temperature $\theta$
has to be replaced by an electron temperature, $\theta_e (t)$,
as yet unknown function of time,
and thus Eq. (8) can be 
reduced to an ordinary differential
equation for $\theta_e (t)$,
where the CMB temperature, $\theta (t)$,
could still be an arbitrary function of time:
\vspace {-.15in}
\begin{equation}
d \theta_e / dt = - 2 \theta^4 (t) 
[ \theta_e (t) - \theta (t) ] / t_C
\tag{10}
\end{equation}
Eqs. (9) and (10) can now be used
to solve the dynamics of both
$\theta = k_B T / m_0 c^2$ and $\theta_e = k_B T_e / m_0 c^2$. 
It suffices, however, to find
$T_e$ as function of $T$;
eliminating the time $t$ 
by dividing Eq. (10) by (9), we get then 
a single equation in the phase space
of $\Theta = T / T_0$, $\Theta_e = T_e / T_0$ as:
\vspace {-.2in}
\begin{equation}
{d \Theta_e} / {d \Theta} =
{\chi \Theta^3  ( \Theta_e - \Theta )} 
( \Omega_{\Lambda} + \Omega_M \Theta^3 )^{-1/2}
\tag{11}
\end{equation}
where we 
dropped the term $\Omega_R \Theta^4$,
which is negligible at $\Theta \lesssim 10^2$, and
introduced a ``QM+cosmic'' parameter
\vspace {-.05in}
\begin{equation}
\chi = 2 ( k_B T_0 / m_0 c^2 ) ^4 / t_C H_0 \approx 1.21 \times 10^{-2} 
 \  ( \approx 5 \alpha / 3  )
\tag{12}
\end{equation}
Note that within known
precision of $H_0$, $\chi$ is well
approximated by $5 \alpha / 3$;
it would be surprising and revealing 
if that is not a chance coincidence.
The boundary condition for the solution
of Eq. (11) is $( \Theta_e - \Theta ) \rightarrow 0$
at $\Theta \rightarrow \infty$.

The numerical solutions 
of Eq. (11) are depicted at Fig. 1.
They clearly show that below $\Theta \approx 20$,
CMB has ``dropped the ball'' and cannot
enforce thermal equilibrium on
cosmic electrons, whose temperature
got eventually frozen at some non-equilibrium level
($T_{\infty} = 16.3 K$ for $\Omega_{\Lambda} = 0.7$) 
till ``the end of time''.
This brings up a new facet to the issue of
``heat death'' of the universe. 
[Note, however, that by our definition of
the density number $\rho^{(e)}$ 
the spacing between
electrons increases as $\Theta^{-1} (t)$.]
This frozen state is fully developed by the present day,
regardless of specific values of $\Omega$'s in Eqs. (9) or (11).
A good analytical approximation for $T_{\infty}$
and the solution of Eq. (11) for various 
$\Omega$'s is found as [14]
\vspace {-.1in}
\begin{equation}
\Theta_e =
\left( \Theta^{5/2} + \Theta_{\infty}^{5/2} \right)^{2/5} \ \ \ \ with
\tag{13}
\end{equation}
\vspace {-.25in}
\begin{equation}
\Theta_{\infty} \equiv
\frac {T_{\infty}} {T_0} = 
\left( \frac {15} 8 \frac {\sqrt{ \Omega_{M} +  3 \chi}} {\chi} \right)^{2/5} 
\tag{14}
\vspace {-.03in}
\end{equation}
Thus conceivable measurements of $T_{\infty}$ in deep space
may offer an alternative way to evaluate
$ \Omega_{\Lambda} \approx 1 - \Omega_{M}$.
\begin{figure}
\includegraphics[angle=270,width=3in]{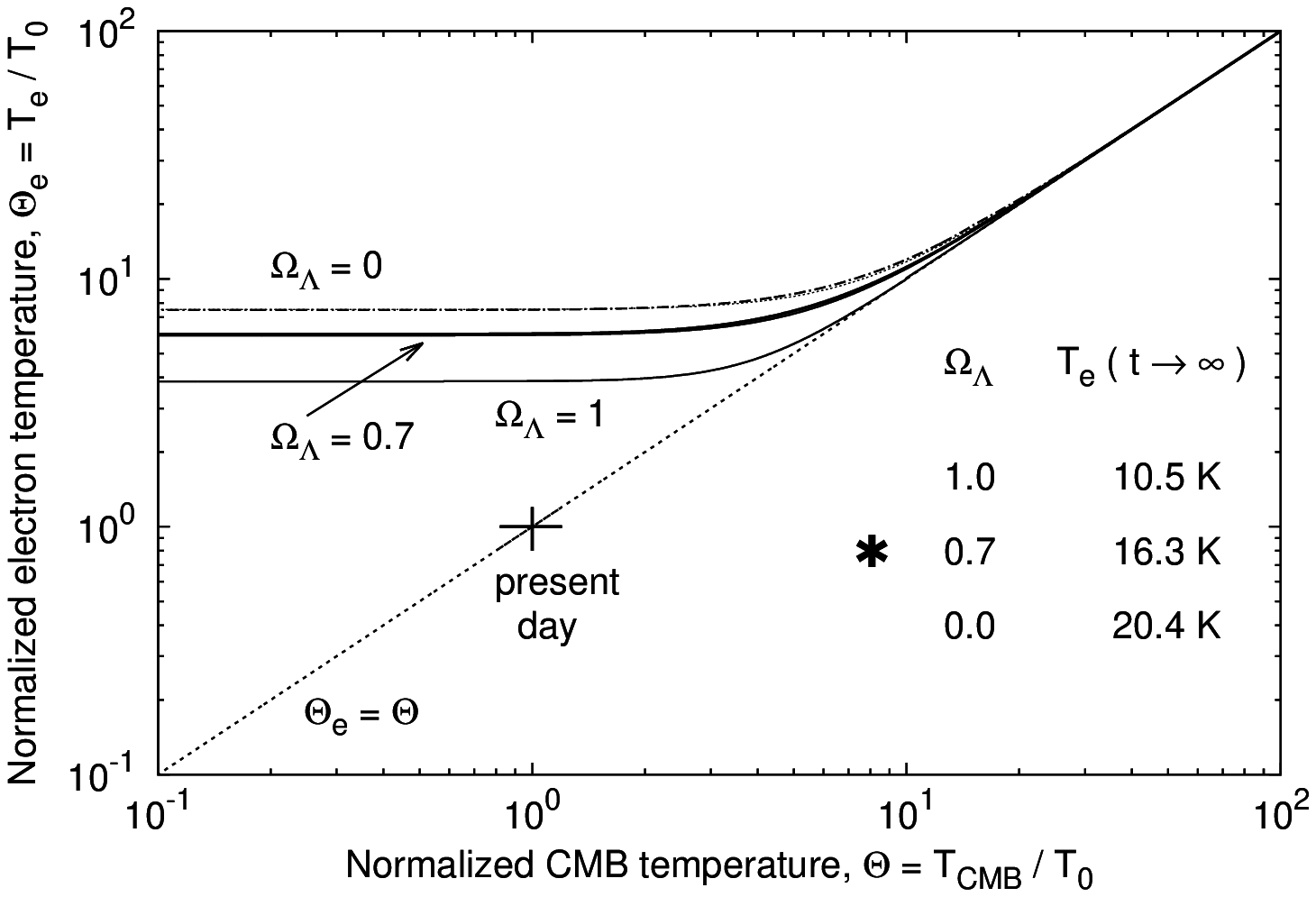}
\caption
{Normalized temperature of electrons $T_e / T_0$,
$vs$ that of CMB, $T / T_0$,
for various values of cosmological constant, $\Omega_{\Lambda}$,
and asymptotic electron temperatures, $T_e$, at $t \rightarrow \infty$.
A star marks the data for a commonly accepted model [11,12].} 
\label{fig1}
\vspace {-.1in}
\end{figure}

{\bf {\emph {Narrow lines and cutoff in cosmic electron spectra}}}?
At the opposite, high-energy end of electron spectrum,
it could be expected that, similarly to baryons,
CMB might strongly affect it,
albeit due to different mechanism,
and do it on a much faster time-scale,
so we can even assume $\theta = const = \theta_0 \ll 1$.
To illustrate that, we consider the dynamics
of the momentum $\mu (t)$ whose implicit solution
for a given $\theta$, is provided by 
$t ( \mu ) = t_C \int d \mu / f ( \mu , \theta )$, Eq. (2).
The integration here can be done numerically,
yet to gain the insights provided by 
analytical results, it would be nice to have a 
``good'' model function $f_M$ that is very close to the
one in Eq. (2) in the domain of interest,
and at that has (a) an analytical integrability of $\int d \mu / f $, 
and (b) explicit ``reversibility'' of resulting functions
$t ( \mu )$ $ \leftrightarrow$ $\mu ( \tau )$.
For $K_C \ll 1$,
Eq. (4) satisfies these conditions and is fully solvable [7].
But to cover both the upper (and largest) part of Thompson domain,
$\mu \gg 1$, and at the same time -- 
the entire immensely larger, Compton  domain, $K_C > 1$,
another greatly useful 
interpolation model is found as
\vspace{-.1in}
\begin{equation}
f_{_M} ( \mu , \theta )  =
- (\theta^3 / q )
y ( 1 + \ln y ) \left\{ 1 +
[ \ln ( 1 + \ln y ) ]^{-2} \right\}^{-1} \ \
\tag{15}
\end{equation}
where $y = 1 + \mu / \mu_C $
with $\mu_C = 1 / q \theta \gg 1$; $\mu >1$.
For $\theta = \theta_0$,
we have $| f - f_M | / f < 0.01 $, for any $\mu > 7$.
At $1 \ll \mu  \ll \mu_C $, 
Eq. (15) yields $f_M \approx - \theta^4 \mu^2 $,
which is consistent with Eq. (3) at $\mu \approx \gamma \gg 1$,
i. e. only for relativistic case.
Yet this is more than enough if $\theta \ll 1$
by insuring that momentum decay
can be continually traced from far Compton
to low  Thompson domains.
Thus Eqs. (4) and (15) smoothly cover the entire
span $\mu \in ( 0 , \mu_{_{Pl}} )$,
as their areas of validity overlap 
by orders of magnitude in $\mu$ if $\theta \ll 1$.
The momentum decay from initial
$\mu = \mu_{in}$ at $\tau = 0$, 
$via$ integration
$d \mu / d t = f_M / t_C$, $\theta = const$ is
\vspace{-.05in}
\begin{equation}
\frac {\mu ( {\tau} )} {\mu_{_C}} =
\exp \left\{ \exp \left[
\sqrt{\frac {( {\tau_0} + {{\tau}})^2} 4 + 1} -
\frac { {\tau_0} + {\tau}  } 2 
\right] -1 \right\} - 1
\tag{16}
\vspace{-.01in}
\end{equation}
where ${\tau} = ({\theta^3}/q) t / t_C$ and
\begin{equation}
{\tau_0} = 
s^{-1} - s \ \ \ with \ \ \
s = \ln [ 1 + \ln ( 1 + 
{\mu_{in}} / {\mu_{_C}} ) ]
\tag{17}
\end{equation}
\begin{figure}
\includegraphics[angle=270,width=3in]{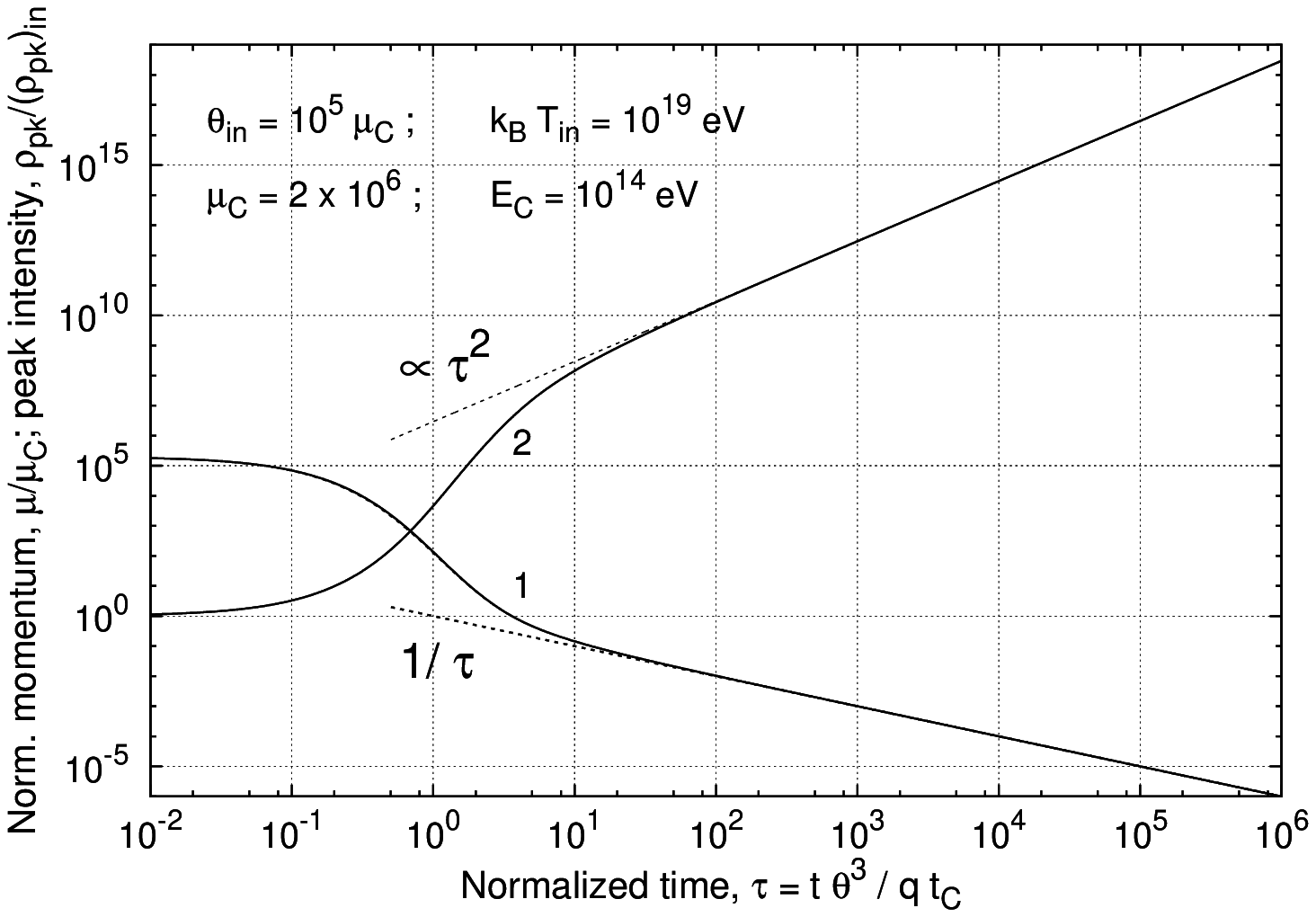}
\caption
{Normalized momentum, $\mu / \mu_{_C}$,
and position of the peak of density distribution (curve 1),
and the peak intensity of the distribution,
$\rho_{pk}/(\rho_{pk})_{in}$, $vs$ normalized
time, ${\tau}$ (curve 2).
Dashed lines -- respective asymptotics 
for ${\tau} \gg 1$. 
}
\label{fig2}
\vspace {-.1in}
\end{figure}
%
For $K_C \approx \mu_{in} / \mu_{_C} = 2 \times 10^5$
(or initial energy 
$E_{in} \sim 2 \times 10^{19} eV$ slightly below
the highest particle energy
$\sim 5 \times 10^{19} eV$, observed in cosmic rays [15]),
$\mu ({\tau})$ is depicted in Fig. 2, curve 1.
The time $\int_{\mu_{_C}}^\mu d \mu / |f_M|$ for an electron
to lose about $2 \times 10^5$ of its momentum
during the ``Compton phase'' $\mu_{in} \rightarrow \mu_{_C}$
is $\Delta {\tau_{_C}} \sim 2.57$, 
hence $(\Delta t)_{_C} \sim  5.3 \times 10^{11} s$,
which is by 6 orders of magnitude 
shorter than the age of universe (and thus justifies
our assumption of $\theta \approx const$),
whereas immediately after that, within the same period,
$\mu$ loses much less than a factor of magnitude.
(For $E_{in} < 5 \times 10^{19} eV$, 
this time is even shorter.)
As $\mu$ keeps decaying from $\mu_{_C}$
down to a relativistic threshold, $\mu = 1$,
its dynamics slows down tremendously,
down to a frozen non-equilibrium at lower $\mu$.

These results call for the study of
the evolution of electron spectra at the energies
far exceeding that of equilibrium.
At that, 
the last term 
in a Fokker-Planck Eq. (7) can be omitted
since $\theta_0 \ll \theta_{in}$,
so that in terms of number density 
$\rho^{(e)} \propto \mu^2 g^{(e)} $ it can be reduced to
\vspace{-.05in}
\begin{equation}
t_C {\partial \rho^{(e)}} / {\partial t} +
{\partial [ f \rho^{(e)} ] } / {\partial \mu} = 0
\tag{18}
\end{equation}
which is essentially a continuity-like equation.
Again, it is fully integrable, and its general solution is
\vspace{-.05in}
\begin{equation}
\rho^{(e)} = {\Phi ( \xi - t )} / {f ( \mu )} ,~~~with~~~
\xi = t_C \textstyle \int  d \mu / f
\tag{19}
\end{equation}
where $\Phi (x)$ is an arbitrary function of $x$
defined here by initial conditions,
e. g. a MJ-distribution with $\theta_{in}  \gg 1$.
A resulting  analytic solution
for $\rho^{(e)} ( \mu , \tau )$ with $f = f_{_M}$, Eq. (15),
for $\rho^{(e)}$ $vs$ $\mu$ for various 
${\tau} = ( \theta^3 / q ) t / t_C$
is plotted in Fig. 3 for initial temperature, 
$\theta_{in} = 10^5 \mu_{_C}$ or $k_B T_{in} = 10^{19} eV$.
\begin{figure}
\includegraphics[angle=270,width=3in]{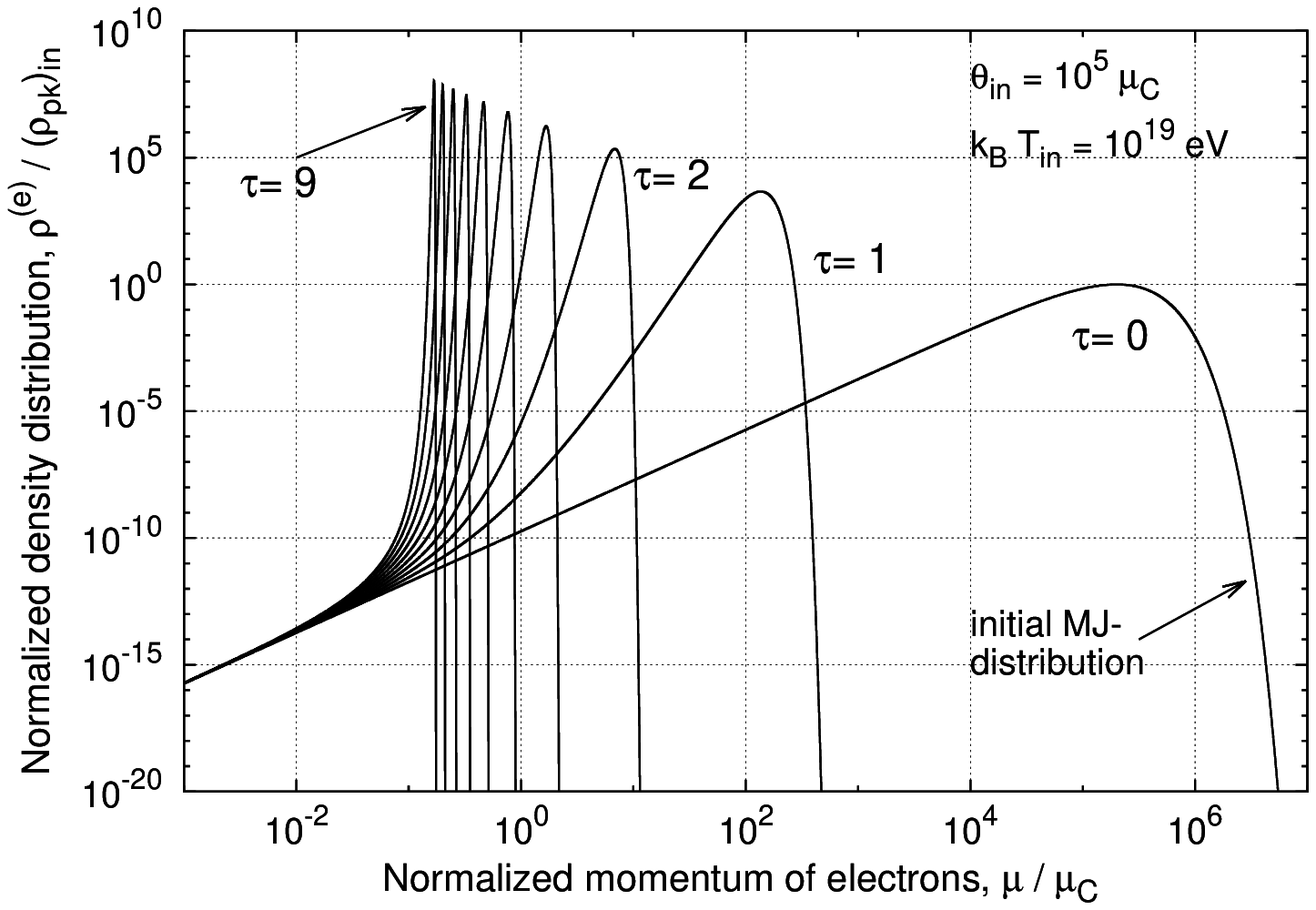}
\caption{
Evolution of normalized density distribution
of electrons, $\rho^{(e)} (\mu ) / (\rho_0^{(e)})_{pk}$
in normalized time, ${\tau}$,
beginning with the initial, Maxwell-J\"{u}ttner (MJ) distribution
at $k_B T_{in} = 10^{19} eV$, from
${\tau} = 0$ to $ {\tau} = 9$
($\Delta {\tau} = 1 \rightarrow 
\Delta t \approx 2 \times 10^{11} s$).
}
\label{fig3}
\vspace {-.1in}
\end{figure}
A curve at ${\tau} = 0$ depicts an initial MJ-distribution,
$\rho_{in}^{(e)} ( \mu ) \propto \mu^2 e^{-\gamma / \theta_{in}}$,
which peaks at $\mu = 2 \theta_{in}$,
i. e. $E_{in} =$ $ 2 \times 10^{19} eV$,
same as for a single electron example.
A transient peak at $\mu = \mu_{pk}$ 
moves fast in the beginning, but slows
down tremendously as it reaches $\mu_{_C} (\theta )$.
Its motion coincides with the timeline of a single electron
with $\mu_{in} = 2 \theta_{in}$, see curve 1 in Fig. 2, 
whereas its intensity $\rho_{pk} ( \tau )$, 
curve 2 in Fig. 2, goes up orders of magnitude
higher than that of the initial MJ-distribution;
at $\mu < \mu_{_C}$, $\rho_{pk} \propto \tau^2$.
Its width narrows down respectively, 
$\Delta \mu / \mu_{pk} \propto \tau^{-2}$
so that for e. g. $\mu = 20$ ($E = 10 MeV$),
it reaches $\Delta E \sim 2 KeV$
i. e.  $\Delta \mu / \mu_{pk} \sim $ $2 \times 10^{-4}$, 
compared to the initial relative width 
$\Delta \mu_{in} / \mu_{in} \sim 1.4 $.
Notice that even before strong line-narrowing,
there is a sharp cutoff at the upper part of the spectrum.
This collapse and cutoff are due to a ``pile-up'' effect,
whereby a leading downward front
moves slower than a trailing one,
resulting in the line squeezing;
it is reminiscent of a shock precursor formation in
astrophysics [16] and Coulomb explosion [17].

These lines would indicate signals 
from far and hot sources;
most likely they will be very weak.
Their detection may necessitate the development of 
high-resolution spectral techniques.
More detailed study 
may need expanding Eqs. (7) and (18) into
anisotropic F-P equations for data analysis.
The averaging over many sources
is expected however to be isotropic,
although the observed line might be broaden up
similarly to the inhomogeneous line broadening
in laser physics [18].
Another major common feature to search for
in these spectra, is a sharp cutoff near
the Compton threshold, $E_C \sim 10^{14} eV$.

In conclusion, we showed that a diminished
light pressure on electrons by CMB
and ensuing low rate of their energy decay should result
in the formation of their frozen non-equilibrium state 
of $ T \sim 10 - 20 K$
as the universe expands long before the current epoch.
We also  predicted the implosion of high-$T$ sources
electron spectra into 
narrow lines and cutoff formation
due to pile-up effect.
\vspace {-.1in}


\begin{thebibliography}{99}
\bibitem{cit13}
In fact, Eq. (8) can be solved the same
way for an $arbitrary$ initial function
$g_{in}^{(e)} (\mu )$,
by decomposing it into exponential components $via$
Laplace transformation,
and looking for the  solutions of Eq. (10) with initial
magnitude ${\theta_e}_{in}$ different for each of the components.
\bibitem{cit14}
If one of $\Omega$'s in Eq. (11)
is unity (hence the other is zero),
Eq. (11) can be integrated in close form in terms of $\Gamma$-function.
In particular, for $\Omega_{\Lambda} = 1$
(and thus the lowest possible temperature $T_{\infty}$),
its exact solution is
$\Theta_e = \Theta + (1/4) (4  / \chi )^{1/4}
exp ( \chi \Theta^4 / 4 ) \Gamma ( 1/4, \chi \Theta^4 / 4 ) $,
where $\Gamma (a,x)$ is an upper incomplete $\Gamma$-function,
and $T_{\infty} / T_0 =$ $ ( 64 \chi )^{-1/4}$ $ \Gamma ( 1/4 ) \approx 3.86$,
where $\Gamma (a) = \Gamma (a,0)$ is a complete $\Gamma$-function.
Thus  $T_{\infty} \approx 10.52 K$, which is consistent
with the numeric calculations, Fig. 1, and Eq. (14).
\end{thebibliography}
\end{document}